\documentclass[twocolumn]{aa} % for final paper version
\usepackage{graphicx}
\usepackage{txfonts}

\def\kms {\ifmmode{{\rm \ts km\ts s}^{-1}}\else{\ts km\ts s$^{-1}$}\fi}
\begin{document}
\title{Chemical and physical small-scale structure in a pre-stellar core}
\titlerunning{Small-scale structure in a pre-stellar core}
\author{Andreas Heithausen\inst{1}\and
        Christoph B\"ottner\inst{2}\and
        Fabian Walter\inst{3}   }
\offprints{A. Heithausen}
\institute{ 
Institut f\"ur integrierte
  Naturwissenschaften, Abteilung Physik, Universit\"at Koblenz-Landau,
  Universit\"atsstr. 1, 56070 Koblenz, Germany,
 \email{heithausen@uni-koblenz.de},
\and
Radioastronomisches Institut, Universit\"at Bonn,
Auf dem H\"ugel 71, 53121 Bonn, Germany,
\email{christoph-boettner@gmx.de}, 
\and
Max-Planck-Institut f\"ur Astronomie, K\"onigstuhl 17, 69117 
Heidelberg, Germany, \email{walter@mpia-hd.mpg.de}
   }
\date{Received 28 August 2007 ; accepted 9 June 2008}

\abstract
{}
{We present a comparative study of several molecular lines and of the dust
contiunuum at 1.2mm in a pre-stellar core that is embedded in the Galactic
cirrus cloud MCLD123.5+24.9.  Previous studies found that the core is
gravitationally stable and shows signs of inward motion. }
{Using the Owens Valley (OVRO) and Plateau de Bure (PdB)
 interferometers we obtained high-angular resolution
maps of the core in the carbon monosulfide 
CS $(2\to1)$ and the cyanoacetylene HC$_3$N $(10\to9)$
transitions. Together with CS $(5\to4)$, C$^{34}$S $(3\to2)$, and bolometer
data obtained with the IRAM 30\,m telescope, we analyse the excitation
conditions and the structural properties of the cloud.}
{On the one hand, the new CS $(J=2\to1)$ observations reveal significant
substructure on a scale of about $7''$, i.e., the beam size,
 corresponding to about 1050 AU at an
adopted distance of 150\,pc. 
On the other hand, the interferometric
observations in the HC$_3$N $(J=10\to9)$ transition shows just one single well
resolved clump in the inner part of the core.  This core is well described by
an intensity profile following from a centrally peaked volume density
distribution. We find no evidence for depletion of CS onto dust grains. 
The inward motion seen in the CS $(2\to1)$ occurs one-sided
from the middle of the filamentary cloud towards the HC$_3$N core.}
{}

\maketitle
%
%________________________________________________________________

\section{Introduction}

Dense cores in molecular clouds are the birthplaces of stars. 
One of the questions of
current interest focusses on the evolution from a core to a protostar. 
Different classifications have been proposed, which are based on different
observational diagnostics. Based on the detection or non-detection of IRAS
point sources, cores have been separated into stellar or starless cores 
(Beichman et al. \cite{beichman86}, Benson \& Myers \cite{benson89}).
Because this classification is limited by the sensitivity and angular
resolution of the IRAS satellite, in recent years some of the cores originally
classified as ``starless'' showed the presence of embedded point sources 
(e.g., Young et al. \cite{young04}).

A different classification scheme was proposed by Ward-Thompson et al.
(\cite{wardthompson94}, \cite{wardthompson99}), who classified cores on the
basis of the non-existence or existence of submm point sources as
pre-protostellar (later on named 'pre-stellar' for brievity, Ward-Thompson et
al. \cite{wardthompson99}) or protostellar.  Unlike the classification of a
core as being protostellar or stellar, 
prestellar cores cover a much wider range in
their evolutionary states. Not all are necessarily gravitationally bound,
while some of them could be in an early stage of contraction. To better
quantify their state it is therefore necessary to study more pre-stellar cores
in different environments.

%-------------------------------------------------------------
\begin{figure*}
\includegraphics[angle=-90,width=17.5cm]{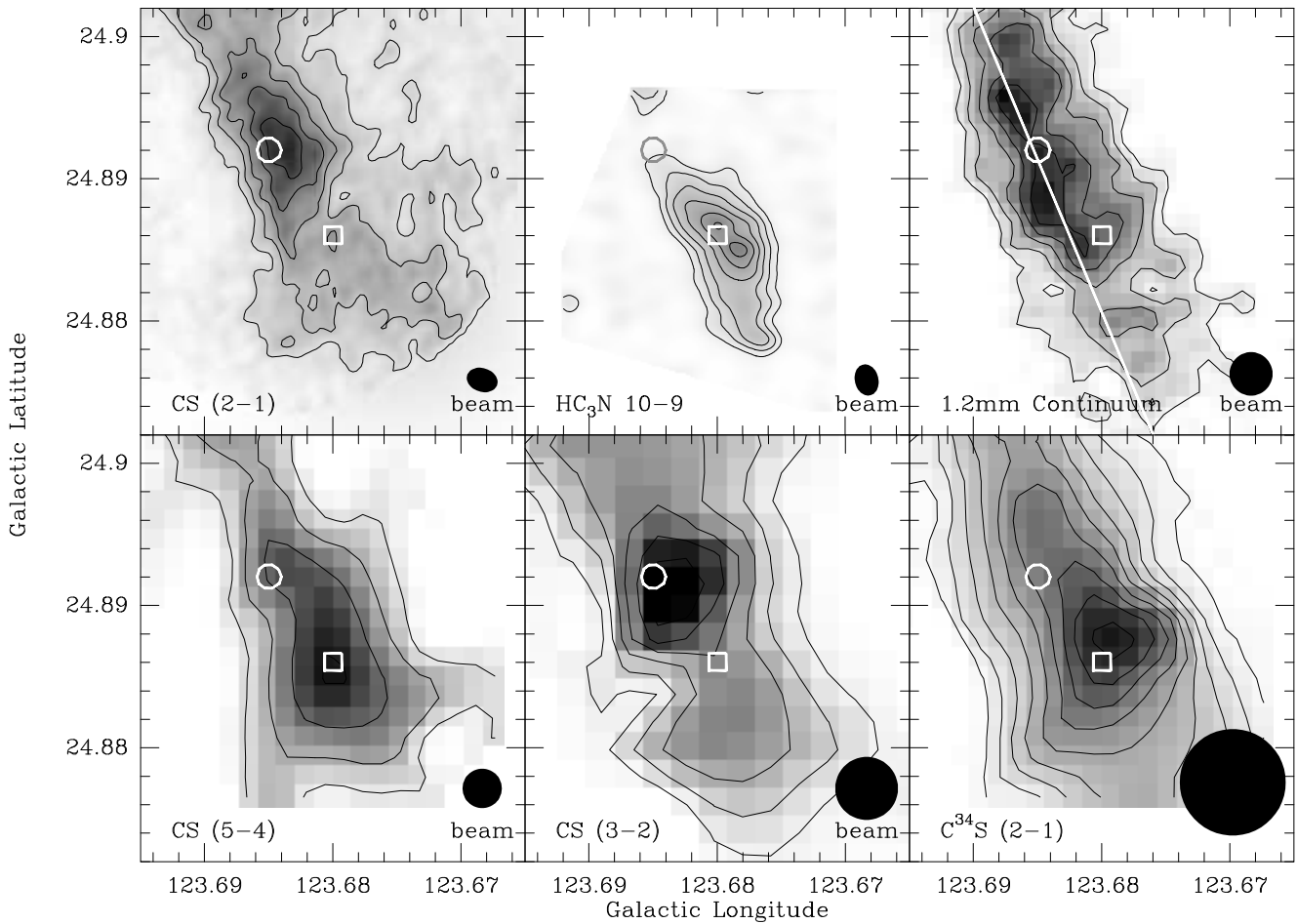}
\caption
{Comparison of integrated intensity maps in different CS lines with maps
in the dust continuum at 1.2mm and in the HC$_3$N $(10\to9)$ line.
 Beamsizes are indicated in the lower
right corner of each map. The open square marks the centre position of
HC$_3$N-B, the open circle that of CS-C.
Contours are every 0.4\,K~km~s$^{-1}$ starting at
0.8\,K~km~s$^{-1}$ for the CS\,$(2\to1)$ line (top left); 
every 0.1\,K~km~s$^{-1}$ starting at
0.2\,K~km~s$^{-1}$ for the HC$_3$N\,$(10\to9)$ line (top middle); 
and every 2\,mJy/beam starting at
2\,mJy/beam for the dust continuum map (top right);
every 0.14\,K~km~s$^{-1}$ starting at
0.14\,K~km~s$^{-1}$ for the CS\,$(5\to4)$ line (bottom left); 
every 0.25\,K~km~s$^{-1}$ starting at
0.25\,K~km~s$^{-1}$ for the CS\,$(3\to2)$ line (bottom middle); and
every 0.06\,K~km~s$^{-1}$ starting at
0.06\,K~km~s$^{-1}$ for the C$^{34}$S $(2\to1)$ line (bottom right).
 The white line in the bolometer map
indicates the major axis of the filamentary cloud.}
\label{integratedmaps}
\end{figure*}
%-------------------------------------------------------------
\begin{figure}
\includegraphics[angle=0,width=9cm]{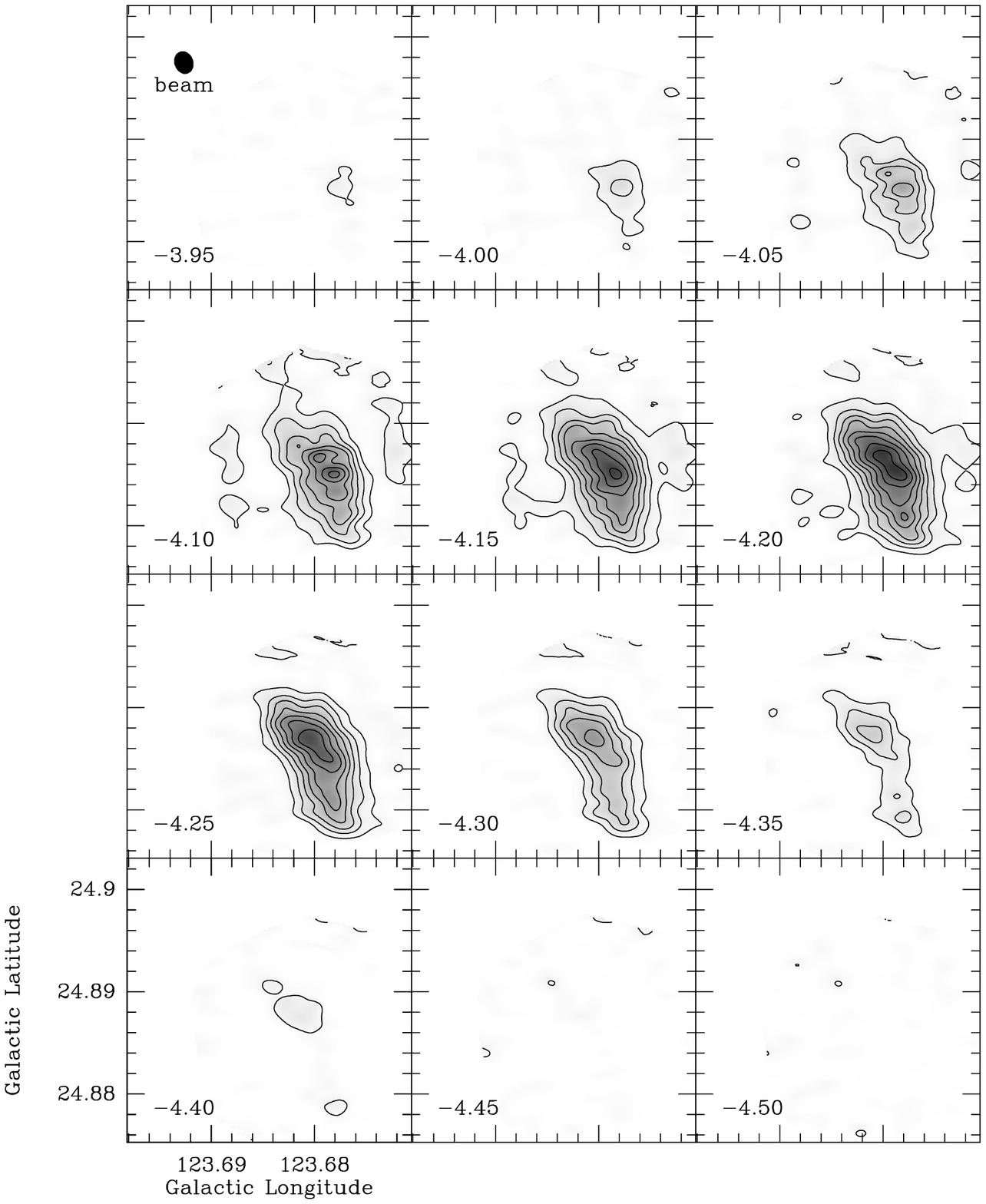}
\caption{Channel maps of HC$_3$N-B in the HC$_3$N\,(10$\to$9) line as obtained
with the Plateau de Bure interferometer.  The beam is indicated in the top
left map. The centre velocity of each channel ($v_{\rm LSR}$) is indicated in
the lower left corner of each map. Contours are every 0.3\,K starting at
0.3\,K.}
\label{hc3nmaps}
\end{figure}
%--------------------------------------------------------------

We continue our study of dense cores in Galactic cirrus clouds
(e.g., B\"ottner \cite{boettner2005}), and concentrate on MCLD123.5+24.9. Based
on the above discussed definitions the core is both starless and
pre-stellar. Nevertheless, CS observations towards part of the core obtained at
high spectral resolution with the IRAM 30m telescope (Heithausen
\cite{heithausen99}, hereafter paper 1) revealed double-peaked CS profiles,
which are interpreted as signature for infall motion (Myers et
al. \cite{myers:etal96}; Choi et al. \cite{choi:etal95}).  This suggested a
scenario of a possibly collapsing fragment embedded in a cloud that is
gravitational unbound on larger scales. Further support for this idea was
given by dust continuum and HC$_3$N spectral line observations, which showed
that the cloud fragment was indeed gravitionally bound (Heithausen et
al. \cite{heithausen02}, hereafter paper 2).

The core itself is chemically evolved, showing a large number of molecules
with abundances similar to other dense cores with on-going star-formation
(Gro{\ss}mann et al. \cite{grossmann90}; Gro{\ss}mann \& Heithausen
\cite{grossmann92}; Heithausen et al. \cite{heithausen95}; Gerin et
al. \cite{gerin97}).  Based on NH$_3$, HCN, HNC, and HCO$^+$ observations, a
 low kinetic temperature of less than 15\,K was derived and its chemical
abundances are well described by low-temperature chemistry with no indication
of shock chemistry (Gro{\ss}mann \& Heithausen \cite{grossmann92}).  The low
temperature was confirmed with observations of the submm- and far-infrared
continuum radiation, which revealed a dust temperature of only 13\,K (Bernard
et al. \cite{bernard99}).  Subsequent observations of SO, CS, CCH, and
C$_3$H$_2$ confirmed high column densities ($N({\rm H_2})\approx
10^{22}$cm$^{-2}$) and volume densities ($n({\rm H}_2)\approx10^5$cm$^{-3}$)
of the core (Heithausen et al.  \cite{heithausen95}; Gerin et
al. \cite{gerin97}).

At an adopted distance of 150pc (Heithausen \& Thaddeus \cite{heithausen90})
the core as seen in the dust continuum at
250GHz or in C$^{18}$O emission has a size of 0.18pc$\times$0.03pc (paper 2).
On that scale the core shows significant variation of its chemical abundances
(Gerin et al. \cite{gerin97}), e.g., cyanoacetylene, HC$_3$N, was found only at
the ends of the filamentary structure, which was interpreted as being caused
by different chemical ages of the different regions (paper 2).

All these results so far were obtained with moderate angular resolution of
single dish telescopes. With such observations it is difficult to get detailed
information on the density structure of the core or to study the connection
between the inward motion and the densest region of the core.  Therefore, 
we present the first high-angular resolution data of this core in
several spectral lines obtained with the Plateau de Bure (PdB) and the Owens
Valley (OVRO) radio interferometers. At the distance of the core we reach an
linear resolution of about 1000\,AU. Details on the observations are presented
in Sect.  \ref{observations}. The results of the observations are discussed in
Sect. \ref{results}. Implications on the density structure and chemistry of
the core are described in Sect. \ref{discussion}. The paper is ended with some
conclusions.

\section{Observations\label{observations}}

The observations of the HC$_3$N\,$(10\to9)$ transition at 90.98 GHz were
conducted between May and November 2002 with the IRAM Plateau de Bure (PdB)
interferometer. During four runs five antennas in the configuration 5D were
used, one run was carried out with 6 antennas in configuration 6Dp. The phase
and amplitude were calibrated with frequent observations of the quasars
1928+738 and 0716+714 and the amplitude scale was derived from measurements of
MWC349 and CRL618. The velocity resolution was 0.05\,km/s channels.

We also observed MCLD\,123.5+24.9 in the CS\,(2-1) line using the Caltech OVRO
millimeter interferometer from January to April 2001. In total, 5 tracks were
spent on source (2 pointings each) in the 'C' and 'L' configurations. The
nearby source J1803+784 (1.5\,Jy) was used for phase calibration.  The
velocity resolution was 0.1\,km/s.

Both interferometric data sets consist of single maps centred on CS-C (paper
1, s. Fig. 1) and HC$_3$N-B (paper 2), 
respectively.  They were corrected for missing
zero spacing using low angular resolution data previously obtained with the
IRAM 30m telescope (paper 1 and 2); for the correction we applied the method
described by Wei\ss\ et al. (\cite{weiss:etal2001}).  In the CS map a larger
fraction of diffuse emission was missing, whereas in the HC$_3$N line most of
the flux was recovered. The effective synthesized beam FWHM
of $6\arcsec \times 8\arcsec$ is identical in both observations however with
different position angles of PA$=-72^{\circ}$ for CS and PA$=-13.5^{\circ}$
for HC$_3$N, respectively; it corresponds to 0.0044pc $\times$\ 0.0058pc or
870AU $\times$\ 1160 AU at the distance of 150\,pc.

The C$^{34}$S $(2\to1)$ and CS $(5\to4)$ observations were observed
simultaneously with the IRAM 30m telescope in September 2000. We mapped the
whole cloud on a 15$''$ grid, which resulted in a total of 82 spectra.  
The beam
size of the telescope was 27$''$ at 96.4\,GHz and 11$''$ at 244.9\,GHz.
Spectra were obtained with an autocorrelation spectrometer set to a velocity
resolution of 0.06\,\kms\ at the lower frequency and 0.048\,\kms\ at the
higher frequency. Main beam efficiencies $\eta_{mb}$ of the telescopes at the
different frequencies were $\eta_{mb}=0.82$ at 96\,GHz and 0.43 at 245\,Ghz.
We use the main beam brightness temperature scale $T_{\rm
mb}$ for the spectral line data.

For the further analysis, we also use our previously obtained CS (3$\to2$) data
(paper 1) and 1.2\ts mm bolometer data (paper 2).

\begin{figure*}
\includegraphics[angle=0,width=16cm]{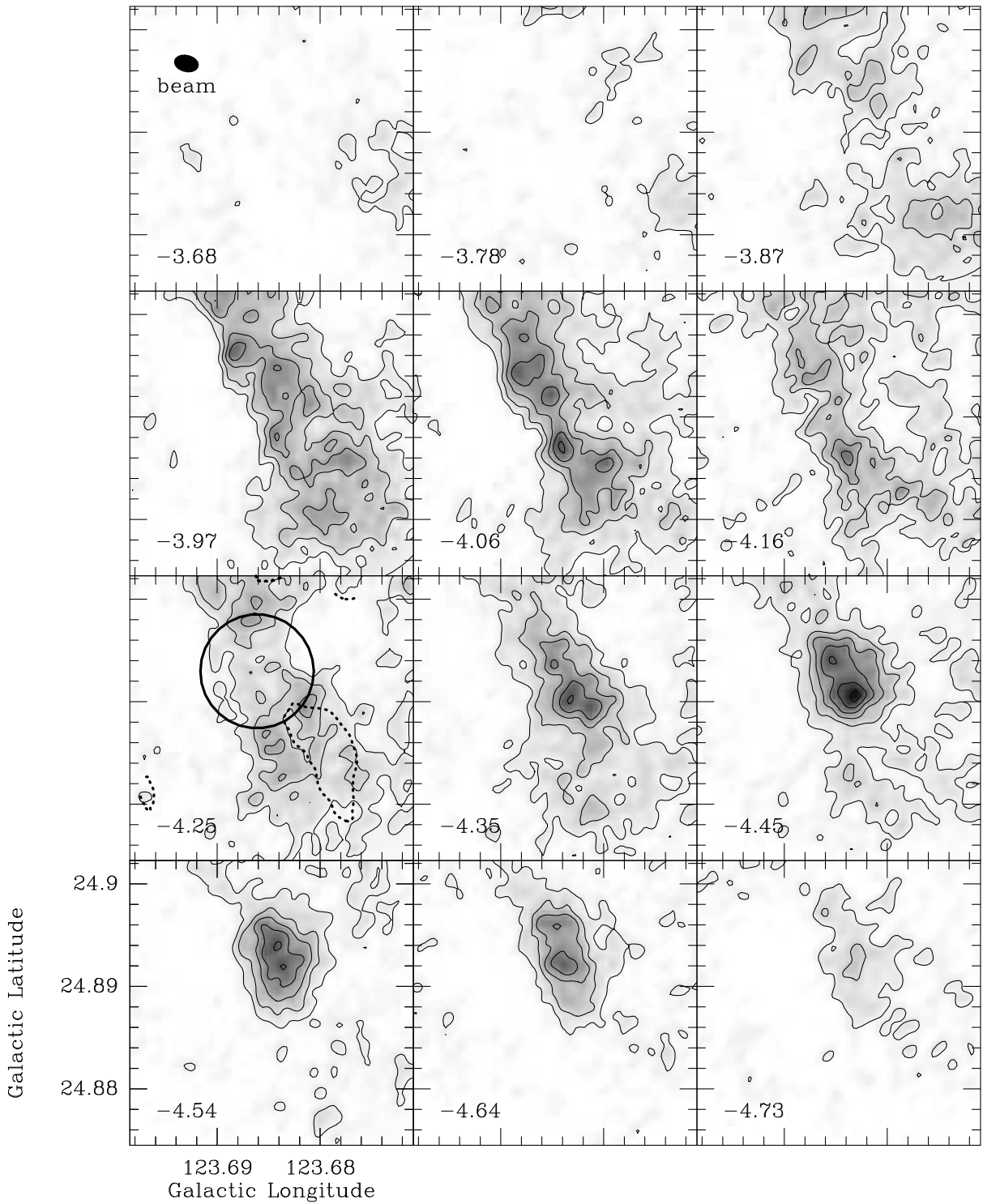}
\caption{Channel maps of CS-C in the CS\,(2$\to$1) line as obtained with the
OVRO interferometer. The centre velocity of each channel $(v_{\rm LSR})$ is
indicated in the lower left corner of each map.  Note that the step between
two channelmaps is larger than in the HC$_3$N maps (s. Fig. \ref{hc3nmaps}).
The beam is indicated in the top left map.  Contours are every 0.75\,K
starting at 0.75\,K. The black circle in the map at -4.25\,\kms\ marks the
region of self-absorption and the dashed contour indicates the position
  and size of the HC$_3$N core.}
\label{csmaps}
\end{figure*}
%--------------------------------------------------------------

\section{Results\label{results}}

\subsection{Comparison of the different data sets}

A comparison of the appearance of the core in different spectral lines and in
the dust continuum is shown in Fig. \ref{integratedmaps}. As was also visible
in previous low angular resolution data (see paper 2), the core has a
filamentary structure that is most pronounced in the dust map.  Taking the
different angular resolutions into account all CS maps look similar, however
with the location of the maximum shifted from higher Galactic latitudes in the
CS $(2\to1)$ and $(3\to2)$ maps towards lower latitudes in the CS $(5\to4)$
and C$^{34}$S $(2\to1)$ maps.

The biggest difference in the appearance of the different molecules is between
the HC$_3$N $(10\to9)$ and the CS $(2\to1)$ maps: the CS emission is spread
over the whole observed region, whereas the HC$_3$N emision is concentrated in
a single clump.  The maximum of the integrated intensity of HC$_3$N (denoted by
HC$_3$N-B in paper 2) is clearly displaced relative to the maximum (denoted by
CS-C in paper 1) of the integrated CS $(2\to1)$ map.  It is, however, found at
the same position as those in the CS $(5\to4)$ and C$^{34}$S $(2\to1)$ maps.

\begin{table}
\caption{Parameters for the HC$_3$N and CS emission identified with GAUSSCLUMP}
\begin{tabular}{l l l l l l l }
\noalign{\hrule}
\noalign{\medskip}
 No. & $l$ & $b$ & 
$T_{\rm mb}$ & $v_{\rm lsr}$ & $\Delta v$ & $\Delta x \times \Delta y$\\
        &  (deg) & (deg)    & 
(K)     & (km/s)    & (km/s)     & ($''\times ''$) \\
\noalign{\medskip}
\noalign{\hrule}
\noalign{\medskip}
\noalign{\hskip 3cm HC$_3$N $(10\to9)$}\noalign{\medskip}
\noalign{\hrule}
\noalign{\medskip}
\#1 & 123.679 & 24.885 & 2.67 & -4.19 & 0.23 & $17.4\times33.2$  \\
\#2 & 123.683 & 24.888 & 0.90 & -4.25 & 0.24 & $ 8.1\times13.0$  \\
\noalign{\medskip}
\noalign{\hrule}
\noalign{\medskip}
\noalign{\hskip 3cm CS $(5\to4)$}
\noalign{\medskip}
\noalign{\hrule}
\noalign{\medskip}
\#1 & 123.680 & 24.886 & 1.3 & -4.09 & 0.36 & $19\times44$  \\
\noalign{\medskip}
\noalign{\hrule}
\noalign{\medskip}
\noalign{\hskip 3cm C$^{34}$S $(2\to1)$}
\noalign{\medskip}
\noalign{\hrule}
\noalign{\medskip}
\#1 & 123.680 & 24.888 & 1.4 & -4.10 & 0.30 & $30\times107$ \\
\noalign{\medskip}
\noalign{\hrule}
\noalign{\medskip}
\noalign{\hskip 3cm CS $(2\to1)$}
\noalign{\medskip}
\noalign{\hrule}
\noalign{\medskip}
\#1 & 123.684 & 24.890 & 4.81 & -4.44 & 0.34 & $ 16.7 \times 35.9$ \\
\#2 & 123.688 & 24.894 & 3.70 & -4.07 & 0.17 & $ 13.2 \times 83.7$ \\
\#3 & 123.685 & 24.895 & 2.45 & -3.97 & 0.21 & $ 10.2 \times 23.5$ \\
\#4 & 123.675 & 24.881 & 2.20 & -3.97 & 0.39 & $ 14.1 \times 31.4$ \\
\#5 & 123.682 & 24.879 & 2.02 & -4.35 & 0.18 & $ 30.7 \times 17.3$ \\
\noalign{\medskip}
\noalign{\hrule}
\noalign{\medskip}
\noalign{Remark: Data are not corrected for finite resolution}
\end{tabular}
\label{clumpstable}
\end{table}

Figures \ref{hc3nmaps} and \ref{csmaps} give an overview of the velocity
structure of the core in the HC$_3$N $(10\to9)$ and CS $(2\to1)$ spectral
lines. The HC$_3$N emission
shows a simple single clump, which is well resolved in our
map. The emission of the clump shows very little
variation from one velocity channel to the next. In contrast to the CS
emission (see below) there is no indication for diffuse HC$_3$N emission.

The CS channel maps show much more structure with several emission maxima and
extensive diffuse emission, which is changing its intensity distribution
significantly from one channel map to the next. The emission is spread over
most of the observed region.

\subsection{Quantification of  the emission}

To derive accurate positions and velocities of the emission throughout the
core, we have quantified the emission seen in the CS and HC$_3$N maps  
using the GAUSSCLUMP algorithm developed by Stutzki \& G\"usten
(\cite{stutzki90}) and discussed by Kramer et al. (\cite{kramer:etal98}). 
This algorithm assumes that the emission of the cloud is composed of
regions that have a Gaussian-shaped intensity distribution. Starting
at the intensity maximum, it fits a single 3-dimensional Gaussian to the
intensity distribution, which is then subtracted. This procedure is repeated
with the residual intensity until the noise level is reached.  In this way, it
subsequently fits the whole cloud into Gaussian-shaped emission blobs.
The results for each spectral line are summarised in Tab. \ref{clumpstable}.

\begin{figure}
\centering
\includegraphics[angle=-90,width=9cm]{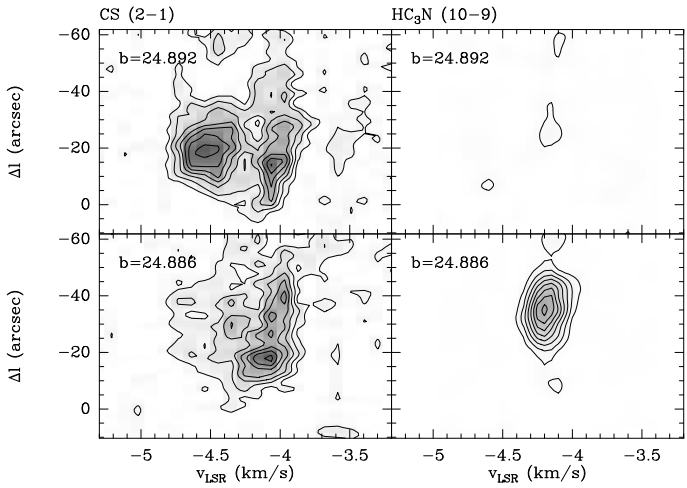}
\caption{Velocity-longitude maps of our high-angular resolution CS
  $(2\to1)$ (left maps) and HC$_3$N $(10\to9)$ (right maps) 
data at the location of the HC$_3$N maximum (lower map) and of the
  CS $(2\to1)$ maximum (upper map). The Galactic latitudes of the cuts are
  indicated at the upper left corner. Contours are every 0.5K starting at
  0.5K for the CS data, and 0.3K starting at 0.3K for the HC$_3$N data. 
Offsets in Galactic longitude are relative to $l=123.\!^\circ689$.} 
\label{lvb-maps}
\end{figure}

\begin{figure}
\centering
\includegraphics[angle=-90,width=9cm]{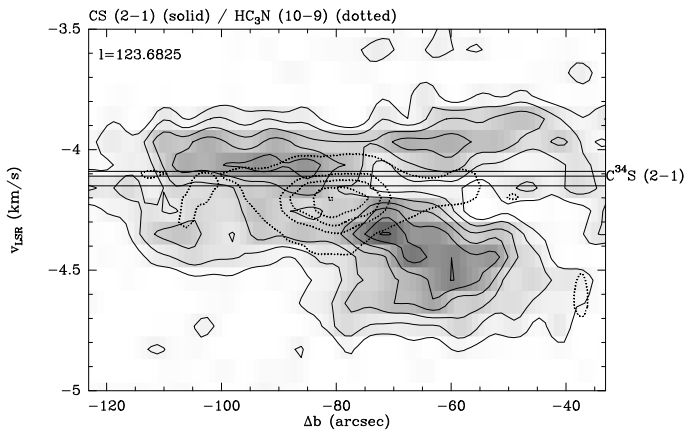}
\caption{Velocity-latitude map of our high-angular resolution CS
  $(2\to1)$ (solid contours) and HC$_3$N $(10\to9)$ (dotted contours) 
data along the Galactic longitude $l=123.\!^\circ6825$. Contours are every 
0.5K starting at 0.5K for the CS data and 0.3K starting at 0.3K for the 
HC$_3$N data. Offsets in Galactic latitude are relative to
$b=24.\!^\circ909$. The solid line at $v_{\rm LSR}=-4.11 {\rm km s^{-1}}$
marks the velocity of the CS $(5\to4)$ and C$^{34}$S $(2\to1)$ lines; the
upper and lower lines parallel to this line indicate the variation of the
centre velocities for these molecules along this cut.}
\label{bvl-maps}
\end{figure}

\begin{figure}
\centering
\includegraphics[angle=-90,width=9cm]{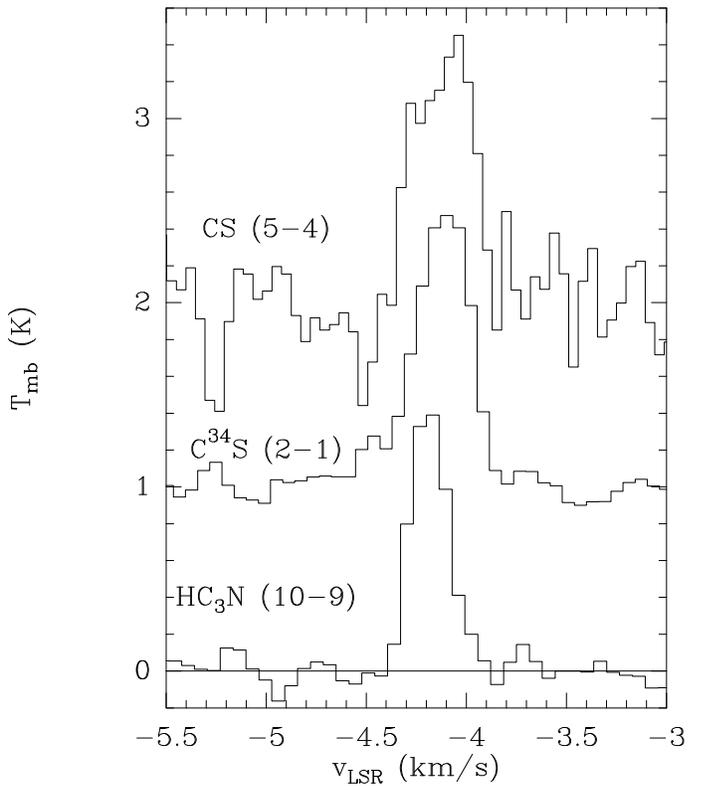}
\caption{Comparison of the CS $(5\to4)$, C$^{34}$S $(2\to1)$, and the
HC$_3$N $(10\to9)$ spectra from the IRAM 30\,m telescope at the position of
the HC$_3$N maximum $(l=123.\!^\circ680, b=24.\!^\circ886)$.  For better
display the upper spectra have been shifted by 1K and 2K, respectively.}
\label{profile-comp}
\end{figure}

The HC$_3$N data are separated into two single clumps. Inspection of the
channel maps by eye (s. Fig. \ref{hc3nmaps}) would probably just define the
emission as a singular clump with a small bending at its lower latitude side.
The computer analysis defines it as two clumps due to the predefined Gaussian
form. More than 90\% of the emission is concentrated in the clumps. 

For comparison with the HC$_3$N data, we have also used the GAUSSCLUMP
algorithm to quantify the CS $(5\to4)$ data. Due to the undersampled maps
(11$''$ beam and 15$''$ sampling) only a single clump was fitted, providing us
the global properties for this transition. The analysis of the C$^{34}$S
  $(2\to1)$ data also provides a single emission region with the same centre in
  position and velocity as the CS $(5\to4)$ line, indicating that both lines
  have similar excitation conditions. 
The results show that the HC$_3$N
clumps have no correspondance in neither the C$^{34}$S $(2\to1)$ nor the CS
$(5\to4)$ 
clumps. While the position of the CS clump on the sky is almost
coincident with HC$_3$N position, it is separated in velocity space by at
least 0.1\,\kms, which corresponds to about half of the line width of only
0.24\,\kms. 

This separation is confirmed by a direct comparison of 
the spectra of the CS $(5\to4)$, C$^{34}$S $(2\to1)$,
and the HC$_3$N $(10\to9)$ lines from the IRAM 30\,m telescope at the
position of the HC$_3$N maximum (Fig. \ref{profile-comp}).
 The Gaussian fit to the spectra provides identical centre velocities for both
 CS lines at $v_{\rm LSR}=-4.11\pm0.01\,\kms$, and identical
line widths of $\Delta v=0.33\pm0.03\kms$, compared to the
HC$_3$N line centred at $v_{\rm LSR}=-4.19\pm0.01$\,\kms, and a much narrower
line width of $\Delta v=0.22\pm 0.02$\,\kms.

For a comparison with the other presumably optically thin lines, we have
also separated the CS $(2\to1)$ emission into Gaussian-shaped regions.  We are
aware that this line is optically thick (see Sect. \ref{displacement}) and that
therefore the identification of these Gaussian-shaped emission regions with
physical substructure in the cloud is not necessarily admissible. We list
however these parameters from our fit, to allow for a better quantitative
comparison with the other spectral lines.  The CS $(2\to1)$ data are separated
into at least 5 individual Gaussian regions, which account for about half of
the total emission, indicating again that a significant fraction of the CS gas
is in a wider spread diffuse component. The maxima of the most intense
($T_{mb}$)\ CS $(2\to1)$ and HC$_3$N clumps are separated by about
25$''$. This analysis also confirms that there is no emission line centred at
the velocity range of the HC$_3$N line. A direct comparison of
velocity-position cuts through our high angular resolution data cubes
illustrates this finding (s. Figs. \ref{lvb-maps} and \ref{bvl-maps}).

\subsection{Comparison of HC$_3$N emission and the infall signature}

The southern most part of the core, CS-C, in MCLD123.5+24.9 has been found to
show asymmetric self-absorbed CS($2\to1$) and ($3\to2$) profiles with the blue
shifted part brighter than the red shifted (paper 1). The velocity of the
($5\to4$) line was coincident with the absorption dip in the lower
transitions. These features are the well tested diagnostics of inward motion
in molecular clouds (Zhou et al. \cite{zhou92}; Walker et al.,
\cite{walker:etal94}; Myers et al. \cite{myers:etal96}).

With our new high-angular resolution data, we are able to study in more detail
the region where the infall signature occurs. The absorption dip can be easily
seen in the channel maps presented in Fig. \ref{csmaps} in the channel at
$v_{\rm LSR}=-4.25$\, \kms. Its position is coincident with the intense CS
emission seen in these maps from the channel at $v_{\rm LSR}=-4.45$\, \kms\
downwards (at $l\approx123^\circ\!.685, b\approx24^\circ\!.892$).
In Fig. \ref{csmaps}, the location of the double-peaked CS
($2\to1$) profiles and that of HC$_3$N-B are marked. 
Position-velocity maps of this
region are shown in Figs. \ref{lvb-maps} and \ref{bvl-maps}.  The infall
signature is positionally adjacent, but not coincident with HC$_3$N-B. Possible
explanations for this displacement will be discussed in
Sect. \ref{displacement}.

\section{Discussion\label{discussion}}

\subsection{Intensity and density profile of the dust filament}

\begin{figure}
\centering
\includegraphics[angle=-90,width=8cm]{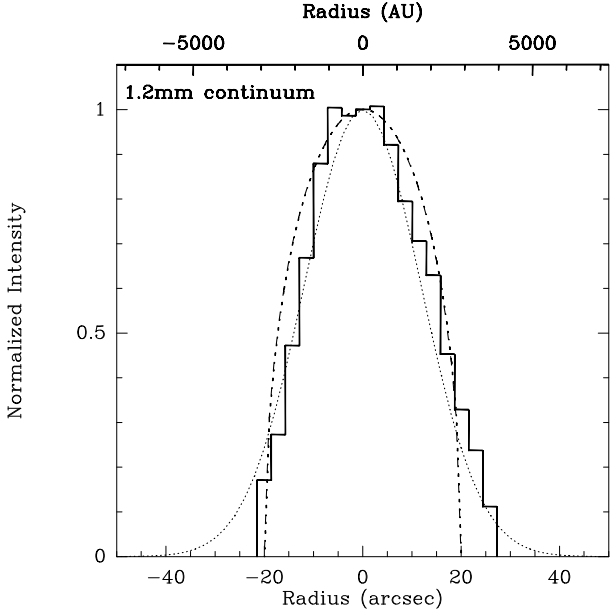}
\caption{Density profile along the minor axis of the dust filament (solid
line) derived by averaging the intensity along the major axis of the
filament. The dotted line represents a Gaussian intensity distribution with a
full-width at half-maximum of 28$''$.  The dot-dashed line corresponds to a
cylinder seen edge-on with constant volume density and a radius of 20$''$.
The upper scale gives the radius in AU, which is derived by adopting a distance
of 150\,pc to the cloud.}
\label{dustprofile}
\end{figure}

In Fig. \ref{dustprofile}, we present an intensity profile of the dust filament
seen in Fig. \ref{integratedmaps}.  The major axis of this filament has a
position angle of $23^\circ$.  We derived the profile by averaging the
intensity along the major axis as function of the minor axis.  The intensity
distribution is well described by a Gaussian profile with a full-width at 
half-maximum of 28$''$. An intensity profile that follows from a constant volume
density assuming optically thin continuum emission does not represent the
observed data as well. A more detailed analysis of the profile is hampered by
the observing technique with which the bolometer data were observed. Due to
the chopping secondary mirror with a chop-throw of 50$''$ to 70$''$, extended
emission on that scale is missing and thus the profile on that scale is
unreliable.

%--------------------------------------------------------------
%--------------------------------------------------------------

%-------------------------------------------------------------
\begin{figure}
\centering
\includegraphics[angle=-90,width=8cm]{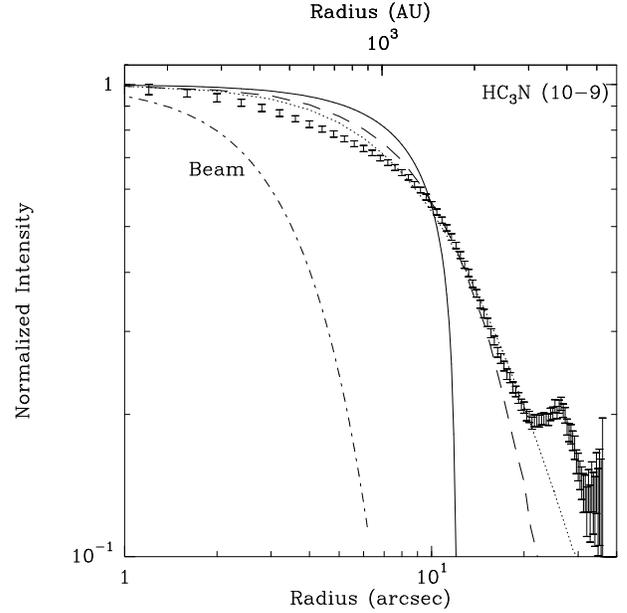}
\caption{Density profile of HC$_3$N-B in the HC$_3$N $(10\to9)$ line (error
bars).  The dot-dashed line corresponds to a beam of 7$''$, which is the
averaged value of the elliptical synthezied beam $(6''\times8'')$.  The
long-dashed line represents a Gaussian intensity distribution with full-widths
at half-maximum of $17''\times 33''$ corresponding to the values listed in
Tab. \ref{clumpstable}. The solid line shows the intensity profile for an
optically thin sphere with constant volume density and a radius of 12$''$.
The dotted line corresponds to a column density profile according to
Eq. \ref{johnstoneprofile} with $\alpha=1.2$ and $r_0=12.''1$.  The upper
scale gives the radius in AU.}
\label{hcn-profile}
\end{figure}
%--------------------------------------------------------------

\subsection{Density profile of the HC$_3$N core}

To determine the intensity profile of HC$_3$N-B we present the circular
average of the intensity distribution centred on HC$_3$N-B of the integrated
$(10\to9)$ line from Fig. \ref{integratedmaps}. Based on the ratios of
the hyperfine components of the lower HC$_3$N spectral lines we have found in
paper 2 that these lines have low optical depths. We derived a volume density
of about $n_{\rm H_2} \approx 3 \times 10^5$cm$^{-3}$. We therefore adopt here
also optically thin emission in the $(10\to9)$ line, i.e., the observed
intensity is proportional to the column density.  In Fig. \ref{hcn-profile}, we
present the results of our profile analysis.  For comparison with the observed
data, we also present some theoretical profiles:
\begin{itemize}
\item the intensity profile of a 2-dimensional structure described by two
Gaussian profiles with full-width at half-maxima of $17''\times 33''$ 
corresponding to the values listed in Tab. \ref{clumpstable}. 
\item a sphere with constant volume density and a diameter of $24''$,
corresponding to the geometric mean of the full-widths at half-maximum derived
from the GAUSSCLUMP analysis (s. Tab. \ref{clumpstable}). 
\item and an intensity profile following
\begin{equation}
I(r) = {I_0\over(1+r^2/r_0^2)^\alpha}
\label{johnstoneprofile}.  
\end{equation}
(cf. Johnstone et al. \cite{johnstone03}). Assuming optically thin emission,
i.e., $I\propto N$, we have varied the projected radius 
$r_0$ and the exponent $\alpha$ over wide ranges and found a best solution for
$\alpha=1.2\pm0.2$ and $r_0=12.''1\pm0.''8$, corresponding to 1760 AU at the
distance of the cloud.
\end{itemize}
It is obvious that the profile which assumes a constant density throughout the
core does not fit the data well. Density distributions with a centrally
peaked density represent the observed intensity distribution much better.
The Gaussian intensity distribution, which follows from a Gaussian density
distribution, has a gradient that is too flat in the inner part and a too steep
gradient at the out part of the core. The column density distribution
following Eq. \ref{johnstoneprofile} is the best representation of the
observed profile.

\subsection{The displacement between CS and HC$_3$N 
\label{displacement}}

The displacement between CS-C as seen in the CS $(2\to1)$ line and HC$_3$N-B
in our high-angular resolution data is striking; it is now much better visible
than in our lower angular resolution data (paper 2). It occurs both in the
plane of the sky  (s. Fig. \ref{integratedmaps}) and in velocity 
(Fig. \ref{lvb-maps} and \ref{bvl-maps}). We
first discuss the possibility of depletion as a  cause for the displacement
before we discuss the velocity field and its implications on infall motion.

\subsubsection{Depletion}
Displacements between CS and an other molecules, as, e.g., N$_2$H$^+$, 
has been seen towards other
prestellar dense cores (e.g., Lada et al. \cite{lada:etal03}; Tafalla et
al. \cite{tafalla02}, \cite{tafalla04}). For the prototypical prestellar core
B68 it has been interpreted as being caused by depletion of CS onto dust
grains (e.g., Tafalla et al. \cite{tafalla02}; Di Francesco et
al. \cite{difrancesco03}).  To study whether
 the explanation, that the displacement
is caused by depletion, holds also for the core under consideration, we
analyse first the optical depth of the CS (2$\to$1) line. The isotopic ratio
of C$^{32}$S/C$^{34}$S in the interstellar medium equals the solar value of
22.7 (Lucas \& Liszt \cite{lucas:liszt98}). Therefore, the intensity ratios
should be equal to that number if both lines are optically thin. The line
averaged ratios, we derive from our observations, are however $6\pm2$ towards
the region with inward motion, corresponding to $\tau_{2\to1}=4\pm1$, and
$2.5\pm0.5$ towards the HC$_3$N peak, corresponding to $\tau_{2\to1}\ge10$. 
This means that the CS $(2\to1)$ lines of both regions have high optical 
depths, with an higher value towards the HC$_3$N core. The
CS (2$\to$1) line is thus not a good tracer for the column density.

There is a much closer agreement between the intensity distributions of the
presumably optically thin transitions of CS $({5\to4})$ and C$^{34}$S
$({2\to1})$ and the HC$_{3}$N $({10\to9})$ line
(s. Fig. \ref{integratedmaps}). This suggests that depletion of CS onto dust
grains does not play a large role for the different distributions.  This is
confirmed by the comparison with our bolometer data: within the limits of the
different angular resolutions there is also a better aggreement between the
optically thin CS$({5\to4})$ and C$^{34}$S lines and the dust emission than
between dust and the CS $(2\to1)$ line (s. Fig. \ref{integratedmaps}).

\subsubsection{Infall motion}

The velocity displacement is obvious in our Gaussian analyis of the
emission. To illustrate this separation we show in Fig. \ref{lvb-maps} two
longitude-velocity maps obtained at different Galactic latitudes through the
core. At the maximum of the CS $(2\to1)$ line there is almost no corresponding
HC$_3$N emission. At this latitude, the self-absorption of the CS line is most
pronounced. Compared to the HC$_3$N line the maximum of the CS line is clearly
blue shifted.  The second position-velocity cut is at the latitude of the
HC$_3$N maximum. At the velocity range of the HC$_3$N line there is only weak
CS emission; the CS profile is assymmetric and the CS maximum is clearly red
shifted compared to the HC$_3$N line.

A latitude-velocity map presented in Fig. \ref{bvl-maps} also illustrates
these findings.  Double-peaked CS profiles are only found towards the maximum
of the CS (2$\to$1) intensity distribution. This region is located at the
northern part of HC$_3$N-B, which indicates the densest part of the core.
Towards the lower latitude side of the HC$_3$N core, the double-peaked CS
lines profiles disappear; only a single peaked asymmetric profile at
redshifted velocities is seen with a broad wing towards the blue side. In this
figure, we also mark the velocity range for the CS $(5\to4)$ and C$^{34}$S
$(2\to1)$ lines, which are clearly redshifted throughout the whole region
compared to the HC$_3$N line.

Double-peaked velocity profiles with the blue component stronger than the red
one are interpreted as being caused by infall motion (Walkers et
al. \cite{walker:etal94}; Choi et al. \cite{choi:etal95}; Myers et
al. \cite{myers:etal96}). Such profiles arise from centrally condensed
collapsing clouds for spectral lines for which the excitation temperature
drops from the core to the outer regions. The formation of such line profiles is
nicely illustrated by Walker et al. (\cite{walker:etal94}) in their
Fig. 1. The near side of the cloud is receeding from the observer, thus
redshifted, while the far side is approaching, thus blueshifted. Due to the
gravitational acceleration by the central core the gas closer to the core has
higher velocities. Therefore, for the redshifted component, the gas with the
lower excitation temperature and the apropriate velocity is between the core
and the observer and thus, if optically thick, can absorb. For the approaching
component the gas with the proper velocity and a lower excitation temperature
is behind the aproaching component and therefore cannot absorb. For spectral
lines with higher optical depth this leads to
the assymetric spectral line profile with a stronger blue than red component.
The velocity of optical thin lines should coincide with the absorption dip.  

The profiles we observe for the upper part of the HC$_3$N core fit 
this description well (s. Fig. \ref{bvl-maps}). The velocity of the HC$_3$N 
coincides with the CS absorption dip. The blueshifted component of the CS
$(2\to1)$ line is stronger than the redshifted. We therefore conclude that the
upper part of the core is indeed contracting.

At the southern or lower part of the core the situation is more complicated:
here the profile reverses; the blueshifted component almost vanishes and thus
the redshifted line is the strongest component.  We note here that the
relation between the redshifted CS component and the HC$_3$N core does not
change much from the upper to the lower part of the core; only the velocity
separation between the CS and HC$_3$N lines decreases. Such a line reversal
has been seen towards other cores, e.g., towards L1544 (Tafalla et
al. \cite{tafalla98}).  One explanation for the reversal is that it is a sign
for outward motion (Walker et al. \cite{walker:etal94}).  If we adopt this
explanation and compare the intensities of the CS lines in the upper and in
the lower part of the core we find that there is more material moving inwards
to the core than outwards from the core.

\section{Conclusions}

We presented a high-angular resolution study of the 
pre-stellar core in MCLD123.5+24.9 in several CS transitions and in the 
HC$_3$N $(10\to 9)$ line as well as in the dust continuum. Our observations 
sharpens  the picture of gravitational collapse in that core. The major 
results of our study are the following:
\begin{itemize}
\item the CS emission shows structure down to the smallest scale oberved,
\item the HC$_3$N emission comes from a single well resolved core,
\item based on an analysis of the intensity profiles in the dust continuum 
and in the HC$_3$N line we find that the core is centrally condensed,
\item towards the upper half of the HC$_3$N core we have clear signs of
  inward motion,
\item towards the lower half the CS profiles could be interpreted as outward
  motion. 
\end{itemize}
The interpretation of the spectral line profiles follows the standard line 
discussed in the literature for
pre-stellar cores. Other possibilities for the interpretation may
exist. Clearly, a radiative transfer modelling is demanded to support or
discard this interpretation.

\begin{acknowledgements}
CB acknowledged financial support from the German Max Planck Society through
the International Max Planck Research School (IMPRS). We thank Carsten Kramer
for critically reading the manuscript. This work was partly supported by the
Deutsche Forschungsgemeinschaft grant SFB-494-C2 and is based on observations
carried out with OVRO and the IRAM telescopes; IRAM is supported by INSU/CNRS
(France), MPG (Germany) and IGN (Spain).
\end{acknowledgements}

\end{document}